\documentclass[a4paper,11pt]{article}
\usepackage{pos}
\usepackage{slashed}
\usepackage{cancel}
\usepackage{braket}
\usepackage{tikz} 
\usepackage{tikz-feynman}

\tikzfeynmanset{compat=1.0.0}

\usetikzlibrary{arrows,shapes}
\usetikzlibrary{trees}
\usetikzlibrary{matrix,arrows} 				
\usetikzlibrary{positioning}				
\usetikzlibrary{calc,through}				
\usetikzlibrary{decorations.pathreplacing}  
\usepackage{pgffor}							

\usetikzlibrary{decorations.pathmorphing}	
\usetikzlibrary{decorations.markings}
\tikzset{
    vector/.style={decorate, decoration={snake}, draw},
	provector/.style={decorate, decoration={snake,amplitude=2.5pt}, draw},
	antivector/.style={decorate, decoration={snake,amplitude=-2.5pt}, draw},
    fermion/.style={draw=black, postaction={decorate},
        decoration={markings,mark=at position .55 with {\arrow[draw=black]{>}}}},
    fermionbar/.style={draw=black, postaction={decorate},
        decoration={markings,mark=at position .55 with {\arrow[draw=black]{<}}}},
    fermionnoarrow/.style={draw=black},
    gluon/.style={decorate, draw=black,
        decoration={coil,amplitude=4pt, segment length=5pt}},
    scalar/.style={dashed,draw=black, postaction={decorate},
        decoration={markings,mark=at position .55 with {\arrow[draw=black]{>}}}},
    scalarbar/.style={dashed,draw=black, postaction={decorate},
        decoration={markings,mark=at position .55 with {\arrow[draw=black]{<}}}},
    scalarnoarrow/.style={dashed,draw=black},
    electron/.style={draw=black, postaction={decorate},
        decoration={markings,mark=at position .55 with {\arrow[draw=black]{>}}}},
	bigvector/.style={decorate, decoration={snake,amplitude=4pt}, draw},
    line/.style={draw=black},
}\usetikzlibrary{decorations.markings}

\title{An example of a bouncing dark matter - EPS-HEP2023}

\author*[a,b]{Bastián Díaz Sáez}

\affiliation[a]{Department of physics, University of Santiago of Chile, Casilla 307, Santiago, Chile}

\affiliation[b]{II. Institut für Theoretische Physik, Universität Hamburg, 22761 Hamburg, Germany}

\emailAdd{bastian.diaz.s@usach.cl}

\abstract{Thermal dark matter, which, after chemical decoupling from the Standard Model thermal plasma rises its yield before it freezes-out, is a new feature of thermal DM scenarios dubbed as "Bouncing Dark Matter". In the following short note, we introduce the topic, exemplifying the mechanism with a simple model that extends the SM with two dark matter particles, a fermion, and a pNGB, plus a second Higgs. We explore the features of the thermal freeze-out of this scenario, with particular emphasis on the exponential growth of the yield of the pNGB. We test the model under collider bounds, relic abundance, direct detection, and we study prospects for indirect detection observables.}

\FullConference{The European Physical Society Conference on High Energy Physics (EPS-HEP2023)\\
 21-25 August 2023\\
Hamburg, Germany\\}

\begin{document}
\maketitle

\section{Introduction}
The connection between the thermal origin of dark matter (DM) and its impact on searches nowadays has become an interesting topic to explore. In particular, when one of the yields of a DM candidate presents a bouncing effect before it freezes-out, i.e., a period of exponential growth for some time, it can have intriguing features in observations today. This manuscript shows the yield behavior and indirect detection (ID) features of a thermal multicomponent DM scenario containing a fermion and a pNGB, supplemented with an unstable second Higgs. Under certain conditions, the pNGB yield in the early universe presents exponential growth for some time, recently dubbed as bouncing DM \cite{Puetter:2022ucx, DiazSaez:2023wli}.


\section{Model}\label{model}
We study the bouncing effect in an extension to the SM which considers two new fields, a singlet Dirac fermion $\psi$ and a complex singlet scalar $\phi$ which acquires vaccum expectation value (vev) $v_s$ (for model details see \cite{DiazSaez:2023wli}). The low energy Lagrangian is given by
\begin{eqnarray}\label{masterlag}
\mathcal{L} \supset - m_\psi\bar\psi\psi - \frac{g_{\psi}}{\sqrt{2}}\bar{\psi}\psi(-h_1 \sin\theta + h_2\cos\theta) - \frac{g_{\psi}}{\sqrt{2}}i\bar{\psi}\gamma^5\psi \chi - V(h_1 , h_2 , \chi),
\end{eqnarray}
where $m_\psi = g_\psi v_s/\sqrt{2}$, $\chi$ is the pNGB with a mass $m_\chi$, and $(h_1,h_2)$ the physical Higgs bosons related to the original ones through $h = \cos\theta h_1 + \sin\theta h_2$ and $s = -\sin\theta h_1 + \cos\theta h_2$, with $\theta$ a mixing angle which satisfies
\begin{eqnarray}\label{thetamix}
 \tan 2\theta = \frac{2\lambda_{HS}v_hv_s}{\lambda_S v_s^2 - \lambda_H v_h^2},
\end{eqnarray}
with $v_h = 246$ GeV. The potential $V(h_1 , h_2 , \chi)$ is shown explicitely in \cite{DiazSaez:2023wli}, which contains the masses and interactions of all the (pseudo)scalars. We identify $h_1$ with the 125 GeV Higgs boson. We choose the free parameters of the model as $\{m_\psi , m_\chi, m_{h_2} ,  g_\psi , \theta \}$. We focus on the case in which $m_\chi < 2m_\psi$, where $\psi$ and $\chi$ are DM candidates simultaneously.


\section{Yields and cross sections}\label{relicabundance}
We assume that in the early Universe both DM candidates were in thermal equilibrium with the SM particles. We solve the evolution of the individual yields $Y_i \equiv n_i/s$, with $i=\psi , \chi$, as a function of the temperature $x \equiv \mu /T$, with $\mu = m_\psi m_\chi /(m_\psi + m_\chi)$. We use \texttt{micrOMEGAs} code \cite{Belanger:2014vza}. 

\subsection{Yields}
\begin{figure}[t!]
\centering
\includegraphics[width=0.32\textwidth]{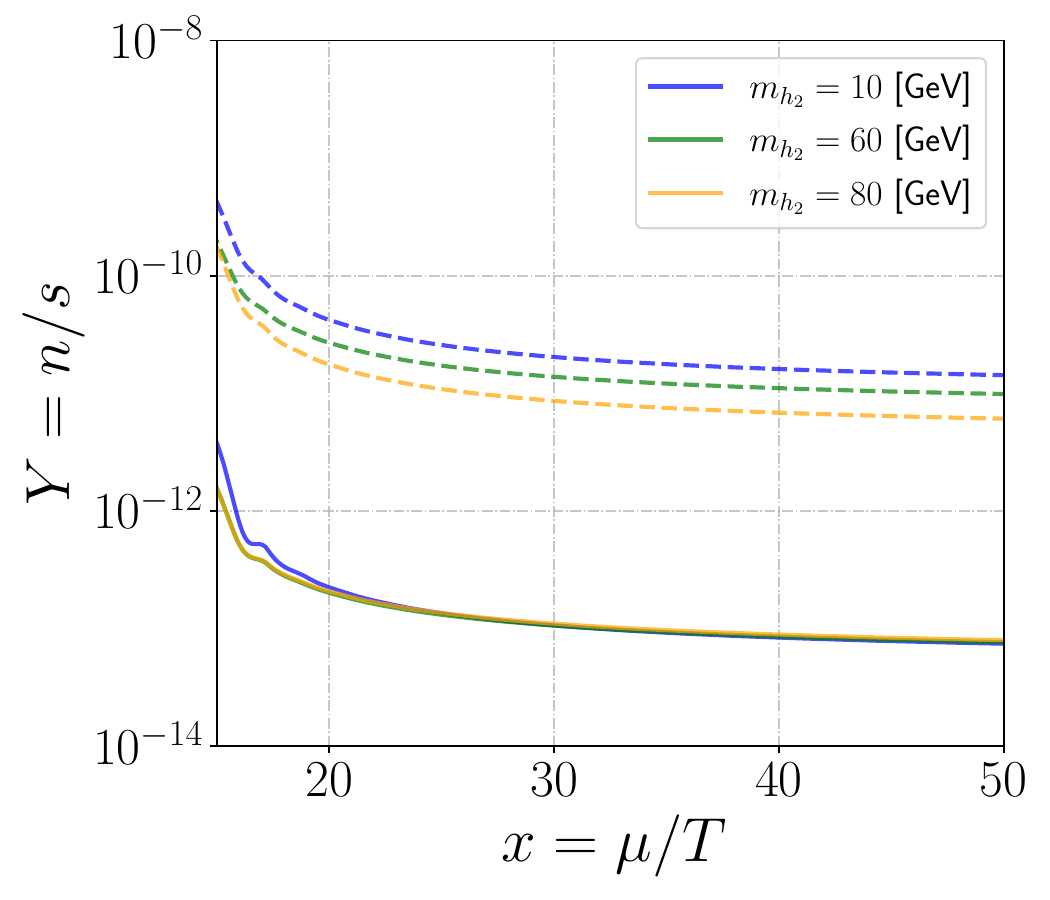}
\includegraphics[width=0.32\textwidth]{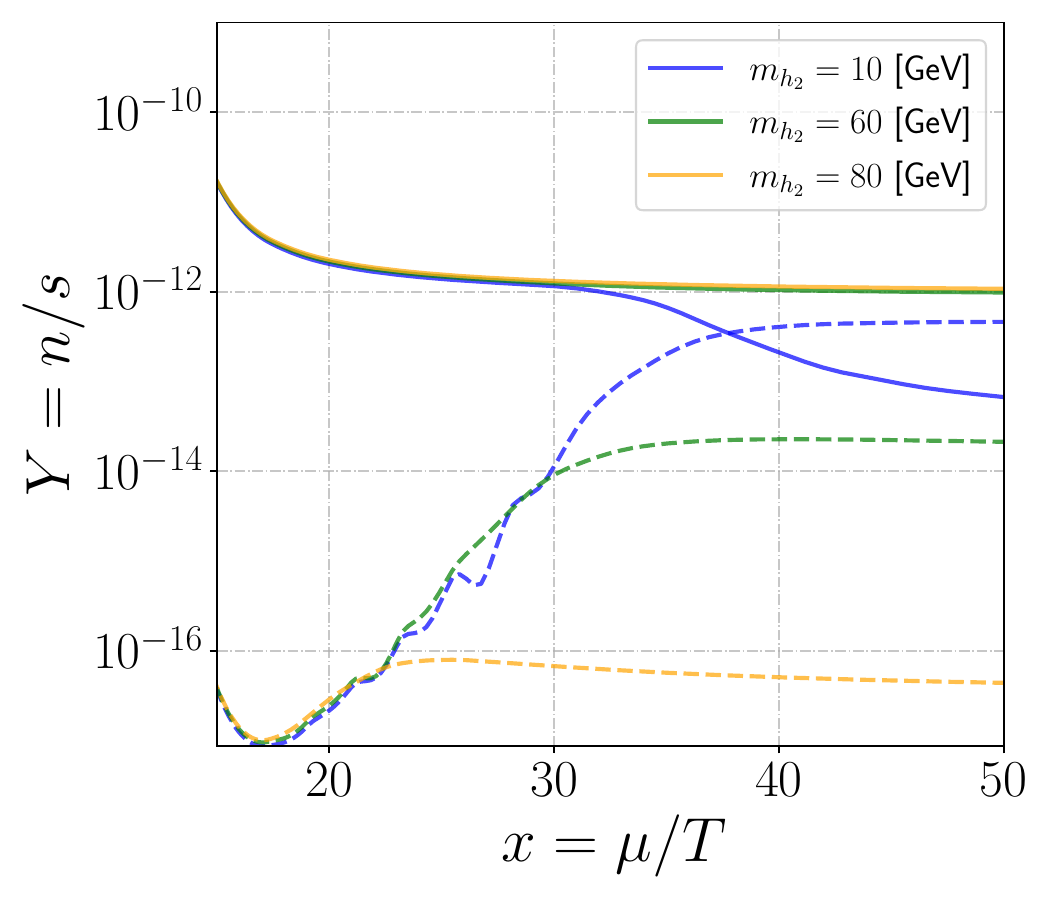}
\includegraphics[width=0.32\textwidth]{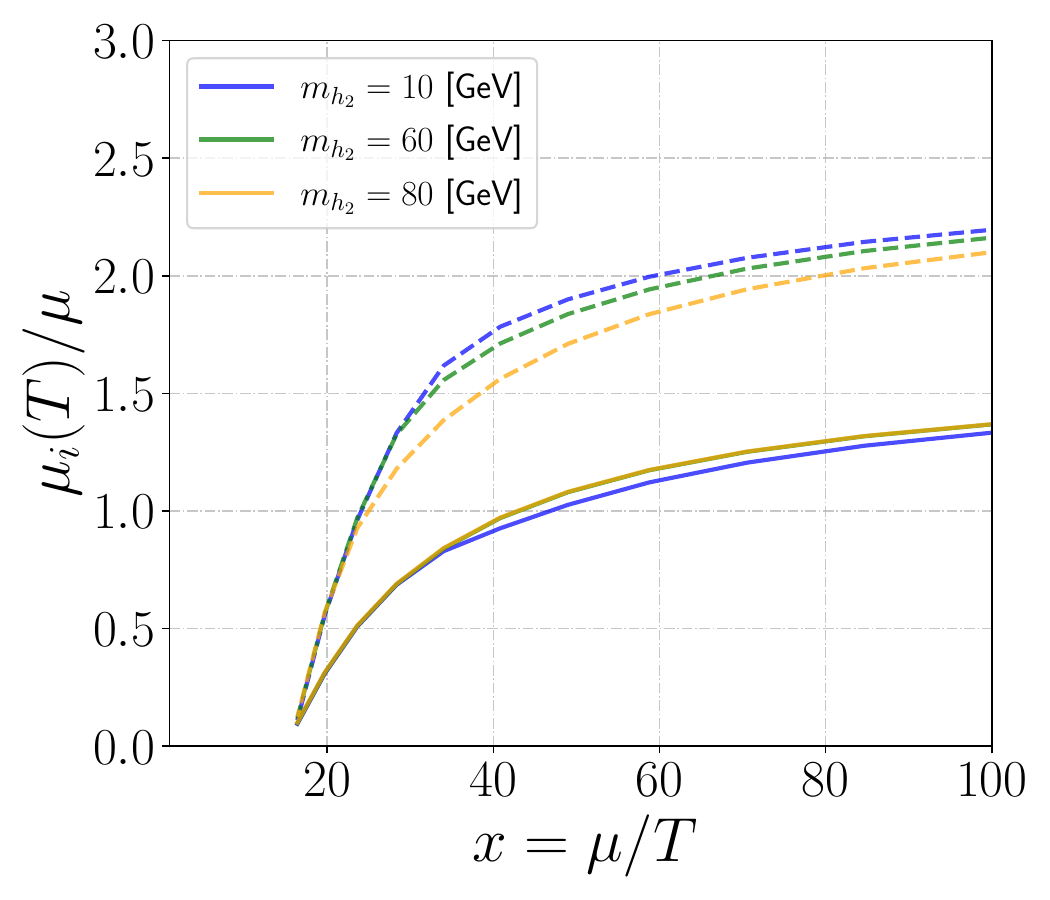}
\caption{(left and middle plots) Yields of $\psi$ (solid lines) and $\chi$ (dashed lines), assuming $g_\psi = 1$ and $\theta = 0.1$. In the plot on the left is considered $(m_\psi, m_\chi) = (200, 150)$ GeV (normal hierarchy), whereas the plot in the middle $(m_\psi, m_\chi) = (100, 150)$ GeV (inverse hierarchy). The plot on the right shows the evolution of the chemical potentials of each DM candidate.}
\label{yields1}
\end{figure}

We distinguish two hierarchies, the \textit{normal hierarchy,} $m_\psi > m_\chi$, and the \textit{inverse hierarchy:} $m_\psi < m_\chi$. In the first one, the freeze-out of the DM particles results to follow the standard freeze-out of two interacting DM particles, each one decoupling from the SM plasma at $x \approx 15 - 20$. We exemplify this with a few parameter space values in Fig.~\ref{yields1}(left).
In the inverse hierarchy at high temperatures, $\psi$ and $\chi$ are in thermal equilibrium with the SM plasma, and both particles have vanishing chemical potentials, $\mu_\psi = \mu_\chi = 0$. Since $m_\chi > m_\psi$, the fermion is less Boltzmann suppressed than $\chi$, and assuming that $\psi$ does not rise in abundance and they remain in chemical equilibrium with each other (same chemical potential), one has that $n_\chi = (n_{\chi,e}/n_{\psi, e})n_\psi \sim e^{-(m_\chi - m_\psi)/T} n_\psi$, i.e., the abundance of $\chi$ decreases as $T$ decreases. However, after chemical decoupling from the SM sector, the DM particles may develop different chemical potentials, entering into a new phase of chemical equilibrium dictated by ractions of the type
\begin{eqnarray}\label{semi}
 \psi + \bar{\psi}\leftrightarrow \chi + h_i, \quad i=1,2.
\end{eqnarray}
These reactions play a key role for some time, increasing the yield of $\chi$ exponentially. This effect is exemplified in Fig.~\ref{yields1}(middle), where the rising of $Y_\chi$ results in a finite temperature interval. The size of the yield increment is model-dependent, in this case depending specifically on $m_{h_2}$. In Fig.~\ref{yields1} (right), we observe graphically the non-zero values taken by the chemical potential of the two stable particles for each value of $m_{h_2}$ as a function of $x$. As $\mu_\psi = \mu_{\bar\psi}$, we have that when the reaction \ref{semi} is in CE, we have that $2\mu_\psi = \mu_\chi$.

\subsection{Known analogies in cosmology}
Two known epochs with similar particle reaction features to this bouncing effect are Big-Bang Nucleosynthesis (BBN) and the Hidrogen formation at the epoch of the decoupling of the CMB. In the former case, one has that at $T\sim \mathcal{O}$(100 keV), for instance, the production of deuterium (D) takes place through
\begin{eqnarray}
 n + p \leftrightarrow D + \gamma ,
\end{eqnarray}
with $n,p,\gamma$ being neutron, proton and photon. As this process occurs in CE, one has $\mu_n + \mu_p = \mu_D$, with $\mu_\gamma = 0$. In that case, deuterium production rises exponentially as $n_D \propto e^{B_D/T}n_p$, with $B_D = m_n + m_p - m_D$. In the case of the CMB, one has that at the recombination epoch, $T\sim \mathcal{O}$(0.1~eV), the production of Hydrogen is triggered by 
\begin{eqnarray}
 e^- + p^+ \leftrightarrow H + \gamma ,
\end{eqnarray}
with this production in CE such that $\mu_e + \mu_p = \mu_H$. Similarly to $D$, the yield of $H$ rises exponentially for some time. Even when these two examples deal with the exponential increase of the yield of bounds states, the exponential growth of $Y_\chi$ behaves similarly.

\subsection{Cross sections}\label{crosssections}
\begin{figure}[t!]
\includegraphics[width=1\textwidth]{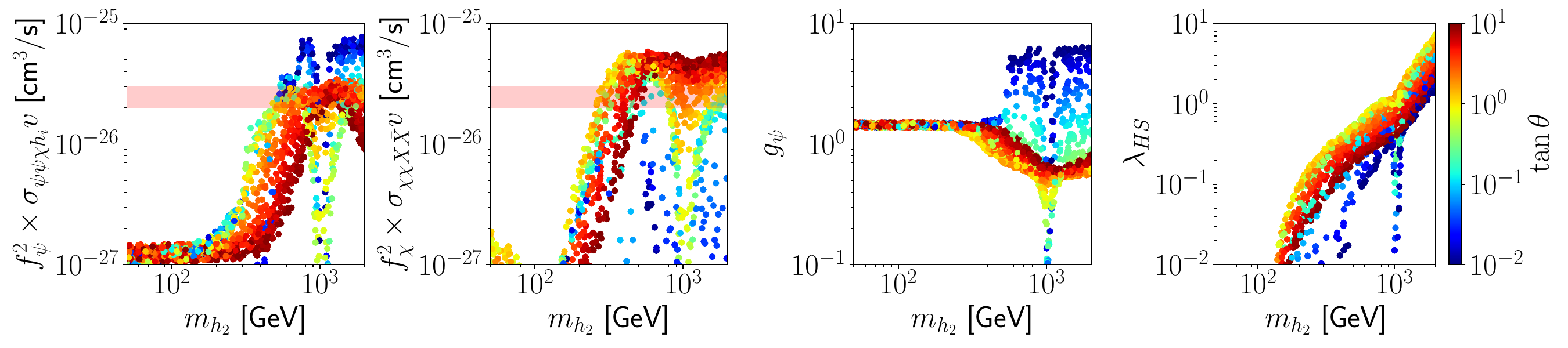}
\caption{Random scan for the parameters $(m_{h_2}, g_\psi, \theta)$ assuming $m_\psi = 500$ GeV and $m_\chi = 700$ GeV, with each point fulfilling the correct relic abundance. In the first two $y$-axis, $f_\text{i} \equiv \Omega_\text{i}/\Omega_c$, with $i = \psi, \chi$.}
\label{random}
\end{figure}

One of the features of this bouncing effect is the enhancement of some cross-sections relevant to ID, keeping the correct relic abundance. In order to quantify this, we show the result of a random scan in the inverse hierarchy considering $m_\psi = 500$ GeV and $m_\chi = 700$ GeV, whereas the rest of the parameters sampled in a logarithmic scale: $m_{h_2} = [50, 2000]$ GeV, $g_\psi = [0.1, 10]$ and $\tan\theta = [10^{-2}, 10^1]$. We have selected all the points that match the observed relic abundance. In Fig.~\ref{random}, we show the results of the random scan in different planes. As seen in the first two plots, many points with weighted cross sections at small velocities surpass the thermal reference value in both the semi-annihilation of the fermion and in the pNGB annihilation into SM particles. In the former case, the highest cross sections are obtained for small $\theta$, then making the channel $\psi\bar{\psi}\rightarrow \chi h_2$ the leading one, whereas, in the case of pNGB annihilations, the highest cross sections require high values of the mixing-angle. We show the corresponding coupling values obtained in the third and fourth plots.

In conclusion, the necessary condition for the bouncing is $m_\chi > m_\psi$, with some weighted cross-sections at low velocities surpassing the canonical thermal cross-section value relevant for ID observables. The partial average cross sections depend strongly on the mixing angle $\theta$.
\section{Phenomenology}\label{pheno}
\subsection{Constraints}
Our perturvative criteria is $\lambda_{HS}\leq \pi$ and $g_\psi\leq \pi$. The most updated Planck result gives today's relic abundance measurement \cite{planck2018}, $\Omega_c h^2 = 0.12$. In our scenario we set $\Omega_c h^2 = \Omega_\psi h^2 + \Omega_\chi h^2$. On the other hand, the direct detection (DD) spin-independent cross section for the pNGB DM candidate vanishes due to its Goldstone nature \cite{Gross:2017dan}. On the contrary, $\psi$ is subject to sizable constraints appearing from the scattering in $t$-channel exchange of $h_1$ and $h_2$. We take bounds from the Lux-Zeplin experiment (LZ) \cite{LZ:2022ufs}. On the other hand, a second Higgs is constrained by two parameters, ($m_{h_2},\theta$). Collider searches set $\theta \lesssim 0.3$ for $m_{h_2} > 100$ GeV \cite{Falkowski:2015iwa}. In what we present in this proceeding, Higgs to invisible constraints do not play any role.

\begin{figure}[t!]
\centering
\includegraphics[width=1\textwidth]{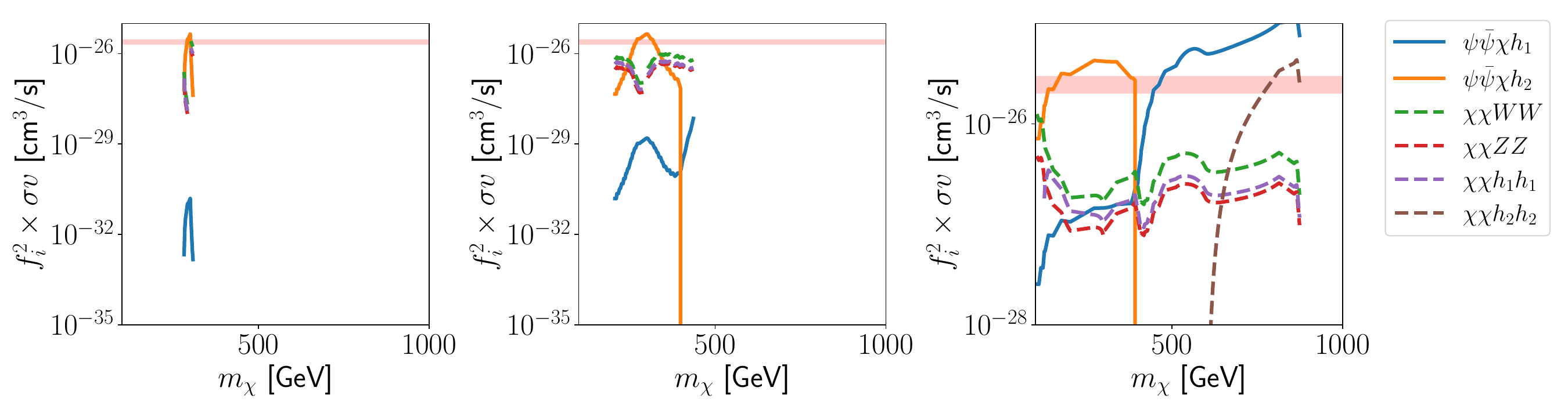}
\includegraphics[width=1\textwidth]{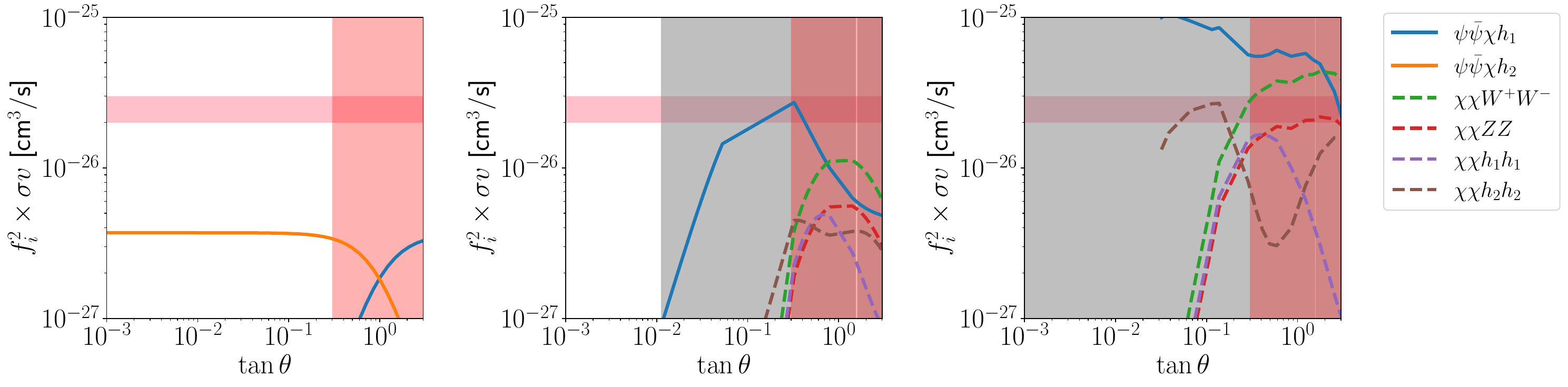}
\caption{(top) Relic weighted average cross section times relative velocity as a function of $m_\chi$. In each plot we have taken $m_\psi = 500$ GeV, $m_{h_2} = 600$ GeV, and $\tan\theta = (10^{-3}, 10^{-2}, 10^{-1})$ (from left to right). We took $g_\psi$ to the correct relic abundance. The pink band represents the thermal canonical value $2-3\times 10^{-26}$ cm$^3/$s. (bottom) Zero-velocity average annihilation cross section for different DM (semi)annihilation channels as a function of the mixing angle. We took $m_\psi = 500$ GeV, $m_\chi = 800$ GeV, $m_{h_2} = (130, 300, 600)$ GeV (from left to right in the plots), and $g_\psi$ takes the necessary value to obtain the correct relic abundance. The pink band references the thermal value, the grey one represents LZ bounds, and the red one collider constraints.}
\label{id1}
\end{figure}

\subsection{Results}\footnote{In the full version of this work \cite{DiazSaez:2023wli}, we explore the viability to have pNGB DM below 50 GeV, founding parameter regions in which this possibility is achieved, contrary to the model in which the fermion is absent at low energies \cite{Gross:2017dan}.}
Considering that the two-DM scenario presents a sizable average annihilation cross section at low temperatures, we focus on (semi)annihilation processes $\psi\bar\psi\rightarrow \chi h_i$ and $\chi\chi\rightarrow XX$, with $X$ an SM state, relevant for ID. The partial cross-sections are highly dependent on the parameters of the model. As we exemplify in the upper row of Fig.~\ref{id1}, considering $m_\psi = 500$ GeV, $m_{h_2} = 600$ GeV and $g_\psi$ getting the appropriate value to match the correct relic abundance, the cross sections not only vary in orders of magnitude depending on $m_\chi$, but as $\theta$ decreases, the parameter space available to obtain the correct relic abundance shrinks allowing only $m_\chi \approx m_{h_2}/2$, otherwise an overabundance is obtained. In this way, in the normal hierarchy and small mixing angles, sizable cross sections can be obtained, as shown by the orange solid line and the dashed curves in the left plot of the upper row of Fig.~\ref{id1}, but only in a reduced parameter space. On the contrary, higher $\tan\theta$ values, e.g., $\tan\theta=0.1$, imply less suppression for (semi)annihilation processes into SM states, including $h_1$ in the final state, presenting strong cross sections, especially in the case $m_\chi > m_\psi$, where the bouncing effect is present. We also can observe this in the third plot of the first row of Fig.~\ref{id1}, showing a wider range of $m_\chi$ allowed. For completeness, we also present the case with $\tan\theta = 10^{-2}$.

We confront the resulting zero-velocity relic weighted cross sections with LZ for points fulfilling the correct relic abundance. In Fig.~\ref{id1}(below), we show the results as a function of the singlet-doublet mixing angle assuming $m_\psi = 500$ GeV, $m_\chi = 800$ GeV, and $m_{h_2} = (130, 300, 600)$ GeV (from left to right, respectively). We show LZ and collider bounds as the color regions. In the left plot of Fig.~\ref{id1}(below), LZ bounds are relaxed due to the algebraic cancellation between the close-in mass of $h_1$ and $h_2$. As $m_{h_2}$ deviates away from $m_{h_1}$, LZ bounds start to be notorious and strong, as it is shown by the middle and the plot in the right, even for $\theta \ll 1$. In this way, all the cross sections with values above the thermal value obtained in the case in which $m_\psi < m_{h_2} < m_\chi$ result disfavored by LZ, although the model still presents sizable ID signatures to be tested by future laboratories.

\section{Conclusions}\label{conclusions}
We have presented a simple extension to the SM, presenting two DM candidates simultaneously. We have found a bouncing yield behavior for the pNGB when this is heavier than the fermion singlet. This model is one of the first scenarios to exemplify the bouncing effect in more detail. We have explored the zero velocity average annihilation cross sections relevant for ID, founding parameter space regions in which both the fermion semi-annihilation and the pNGB annihilation today present values above the canonical thermal value. DD and collider constraints force the mixing angle $\theta$ to be small, e.g., $\theta \lesssim 0.1$, so the strongest signals relevant for ID become disfavored in this scenario. However, the model still presents testable ID signals in the ballpark of future experiments such as CTA.

\section{Acknowledgments}
B.D.S has been founded by ANID (ex CONICYT) Grant No. 3220566. B.D.S also wants to thank DESY and the Cluster of Excellence Quantum Universe.

\bibliographystyle{JHEP}
\bibliography{bibliography}

\end{document}